\documentclass[journal]{IEEEtran}

\ifCLASSINFOpdf
\else
   \usepackage[dvips]{graphicx}
\fi
\usepackage{url}
\usepackage{cite}
\usepackage{bbding}
\usepackage{amsfonts}
\usepackage{graphicx}
\usepackage{eucal}
\usepackage{makecell}
\usepackage{multirow}
\usepackage{array}
\usepackage{mathtools}
\usepackage{booktabs}
\usepackage{tipa}
\usepackage{amsmath}
\usepackage{xcolor}
\usepackage{hyperref}

\begin{document}

\title{{ExPO}: {Ex}plainable {P}honetic Trait-{O}riented Network for Speaker Verification}

\author{Yi Ma, \IEEEmembership{Student Member, IEEE}, Shuai Wang, \IEEEmembership{Member, IEEE}, Tianchi Liu, \IEEEmembership{Student Member, IEEE},
Haizhou Li, \IEEEmembership{Fellow, IEEE}
%\thanks{This paragraph of the first footnote will contain the date on which you submitted your paper for review. It will also contain support information, including sponsor and financial support acknowledgment. For example, ``This work was supported in part by the U.S. Department of Commerce under Grant BS123456.'' }
\thanks{This work is partly supported by  Shenzhen Science and Technology Program (Shenzhen Key Laboratory Grant No. ZDSYS20230626091302006), Shenzhen Science and Technology Research Fund (Fundamental Research Key Project Grant No. JCYJ20220818103001002), China NSFC projects under Grants No. 62401377 and 62271432. We would like to acknowledge that computational work involved in this work is partially supported by National University of Singapore IT’s Research Computing group and Shanghai Jiao Tong University BiCASL Lab. \textit{(Corresponding author: Shuai Wang)}

Yi Ma and Tianchi Liu are with the Department of
Electrical and Computer Engineering, National University of Singapore,
119077, Singapore. (e-mail: mayi@u.nus.edu and tianch\_liu@u.nus.edu) 

Tianchi Liu is also with the Institute
for Infocomm Research, A$^\star$STAR, 138632, Singapore.

Shuai Wang and Haizhou Li are with Shenzhen Research Institute of Big data, School of Data Science, The Chinese University of Hong Kong, Shenzhen 518172,
China; (e-mail: wangshuai@cuhk.edu.cn and haizhouli@cuhk.edu.cn)
}}

\markboth{Journal of \LaTeX\ Class Files, Vol. 14, No. 8, August 2015}
{Shell \MakeLowercase{\textit{et al.}}: Bare Demo of IEEEtran.cls for IEEE Journals}
\maketitle

\begin{abstract}
In speaker verification,  we use computational method to verify if an utterance matches the identity of an enrolled speaker.
%, crucial for applications like forensic speaker comparison. 
%
This task is similar to the manual task of forensic voice comparison, where linguistic analysis is combined with auditory measurements to compare and evaluate voice samples. 
Despite much success, we have yet to develop a speaker verification system that offers explainable results comparable to those from manual forensic voice comparison.
% or aim for intrinsic interpretability, 
%Traditional approaches provide post-hoc explanations for trained models with methods like saliency maps highlighting key features. 
% or self-explanatory models that inherently reveal how features are recognized. 
%
%However, these methods often fall short in open-set scenarios or when different trials focus on varied important regions, thereby limiting practical explanations. 
% 
A novel approach, Explainable Phonetic Trait-Oriented (ExPO) network, is proposed in this paper to introduce the speaker's \textit{phonetic trait} which describes the speaker's characteristics at the phonetic level, resembling what forensic comparison does.  
%, which are physiologically unique and carry speaker-specific characteristics. 
%
%By adding phoneme-level layers to the conventional model, 
ExPO not only generates utterance-level speaker embeddings but also allows for fine-grained analysis and visualization of phonetic traits, offering an explainable speaker verification process. 
Furthermore, we investigate phonetic traits from within-speaker and between-speaker variation perspectives to determine which trait is most effective for speaker verification, marking an important step towards explainable speaker verification. Our code is available at \href{https://github.com/mmmmayi/ExPO}{https://github.com/mmmmayi/ExPO}. 

\end{abstract}

\begin{IEEEkeywords}
speaker verification, speaker phonetic trait, explainability
%Enter key words or phrases in alphabetical order, separated by commas. For a list of suggested keywords, send a blank e-mail to keywords@ieee.org or visit \url{http://www.ieee.org/organizations/pubs/ani_prod/keywrd98.txt}
\end{IEEEkeywords}

\IEEEpeerreviewmaketitle

\vspace{-0.3cm}
\section{Introduction}
\begin{figure}
\centerline{\includegraphics[width=7.5cm]{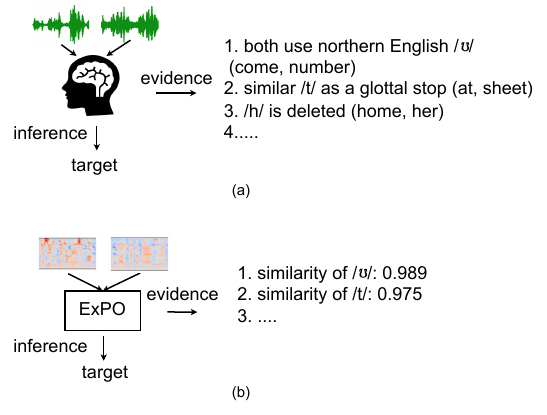}}
\vspace{-0.3cm}
\caption{Illustration of the explainability  in manual voice comparison and neural speaker verification system. (a) a manual voice comparison process that is explainable;
%(b) neural speaker verification that is a black box; (c) prior study that produces a spectrogram heatmap  of the test speech sample as a visual explanation; 
(b) The proposed ExPO model that performs speaker verification in a similar way as manual voice comparison.}
\vspace{-0.6cm}
\label{intro}
\end{figure}

\IEEEPARstart{T}{he} speaker verification task involves determining computationally whether a test utterance matches the voice identity of a target speaker~\cite{bimbot2004tutorial,10447138,liu2021phoneme,10531230,9688243,9413351}. This is similar to the manual task of forensic voice comparison, where we first extract features from the speech of test utterances as well as the target speaker, then make decision by comparing the features~\cite{nolan1983phonetic,morrison2019introduction}, as illustrated in Fig.~\ref{intro} (a). The feature extraction techniques have been intensively studied. The linguistic-auditory method is one of them~\cite{foulkes2012forensic}. This method breaks down an utterance into fine-grained constituent units, such as consonants, vowels, and intonation. The units are then compared by some analytical techniques from the perspective of phonetics, phonology, acoustics, and sociolinguistics. 

Such two-step approach offers two benefits. First, it provides a fact-based reasoning process, that is explainable and trustworthy.  Second, it only pays attention to the most prominent discriminative features, which may lead to a higher accuracy~\cite{wolf1972efficient,pruzansky1964talker}. 
As shown in Fig.~\ref{intro} (a), one evidence is the speaker information contained in phone [\textipa{\textupsilon}], that appears in the word `come' and the word `number'. This feature shows high similarity in that [\textipa{\textupsilon}] in these two words are both typical Northern English accents.

%considering that not all features effectively reflect speaker information, analyzing and selecting more effective features leads to more accurate inferences.
%efficient features, which display small within-speaker variation, but large between-speaker variation, lead to more accurate inferences~\cite{wolf1972efficient,pruzansky1964talker}. 
%features of speaker information are extracted with the help of phonetic transcription. 
%the human speaker comparison task involves observing both test and enrolled utterances on a set of efficient parameters %such as phonemes with small within-speaker differences but large between-speaker differences, 
%and comparing these observations to give the final decision~\cite{foulkes2012forensic}. 

Despite significant progress, neural speaker models are seen as a black box~\cite{desplanques2020ecapa, 8461375,wang23ha_interspeech,thebaud2024phonetic,zhang23aa_interspeech,9053871}. Typically a neural speaker model encodes a spoken utterance into a speaker embedding. By comparing a test speaker embedding and a target speaker embedding, we derive a similarity score between the two embeddings. This process does not explain how the similarity score is derived.
%, i.e. the similarity between a test and a target utterance. %. Fig.~\ref{intro} (b) shows the working process of the deep learning models, where only the final score indicating the level of similarity between the two utterances is given. %Users can't determine what features the model used to make decisions, nor can they determine how effectively these features reflect the characteristics of speakers.
%In spite of its excellent accuracy, this method's black-box nature makes it difficult for users to fully trust the prediction%, which restricts its use in situations such as courts and hospitals.
%As Fig.~\ref{intro} (c) shows, although many 
Some attempt to visualize the decision processing, i.e. with spectrogram heatmap, to explain the decision~\cite{selvaraju2017grad,li22l_interspeech, li2023visualizing, zhang23x_interspeech,10426806}. %However, this is not . For example, a spectrogram heatmap identifies which time-frequency (t-f) bins of a spectrogram the model relies on for decision making~\cite{selvaraju2017grad, 9462463,li22l_interspeech, li2023visualizing, zhang23x_interspeech,yao2023symmetric,10426806}, %it is difficult for a non-expert to understand the physical meaning of these highlighted t-f bins as human voice comparison did. 
%it still falls short of full transparency.
Such heatmap does not really explain the decision in human terms.
Some studies have attempted to analyze speaker attribution during speaker recognition, which is similar to the phonetic trait proposed in our paper~\cite{wu2024explainable,ben2023describing,luu2020leveraging,hong2023decomposition,chen2021phoneme,liu2018speaker,9681187,wang22h_interspeech}. However, these approaches either trade off explainability with a sharp decline in model performance or lack understandable comparative observations between enrollment and test utterances.

In this paper, we propose an explainable phonetic trait-oriented speaker verification model, i.e.  ExPO, as shown in Fig.~\ref{intro} (b). ExPO seeks to perform speaker verification task by comparing the \textit{phonetic trait} across speech samples, which is explainable in human terms. % for the speaker verification task as a way to align the model's decision-making process with human perception and provide comprehensible evidence for laypeople. 
%In th the feature extracted by the linguistic-acoustic method, the \textit{phonetic trait}, which describes the speaker information contained in each phoneme, is introduced in our model.

The rest of this paper is organized as follows. In Section II, we describe the architecture of the phonetic traits-oriented network. In Section~\ref{sec3}, we formulate the training process. The experimental setup, dataset and evaluation methods are presented in Section~\ref{sec4}. Section~\ref{sec5} shows our experimental results. Finally Section~\ref{sec6} concludes the paper.

\vspace{-0.2cm}
\section{Speaker model with Phonetic Trait}
\label{sec2}
\begin{figure}[t]
  \centering
  \includegraphics[trim=0cm 0.7cm 0cm 0cm,width=\linewidth]{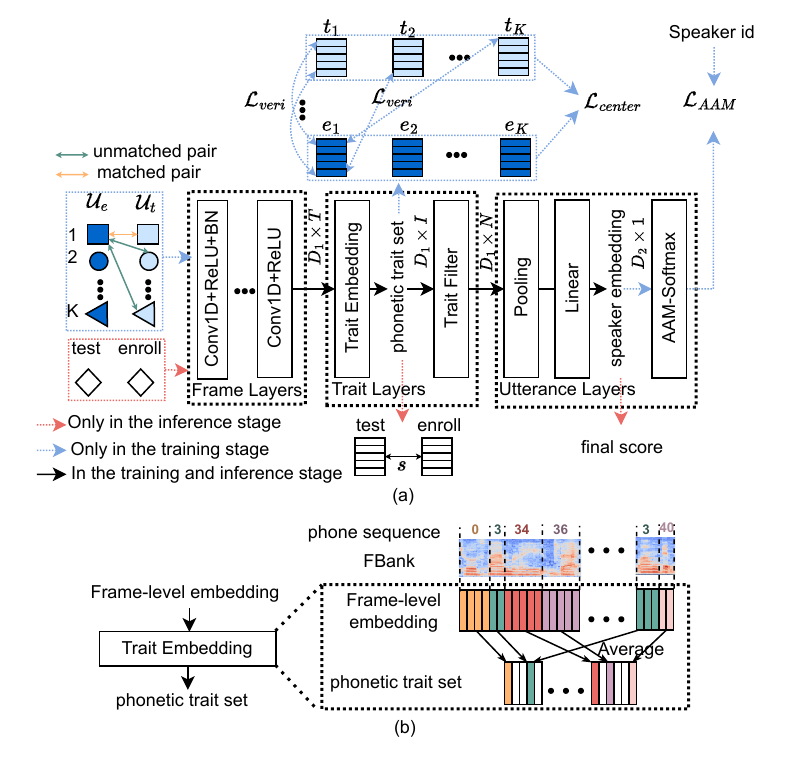}
  \vspace{-0.5cm}
  \caption{Block diagram of the proposed ExPO system.
(a) ExPO introduces trait layers between frame layers and utterance layers of a standard ECAPA-TDNN network. During training, data are sampled pairwise as inputs, with the same shape indicating utterances from the same speaker. %\textcolor{red}{Eucl. is the Euclidean distance.???} 
At the inference stage, the system generates the final score and phonetic similarity vector from the speaker embedding and phonetic traits, respectively. (b) The process in a trait embedding layer. }
\vspace{-0.3cm}
\label{fig:architecture}
\end{figure}

% \textcolor{red}{(please explain what is the black-box model, and justify why we use it as the base to insert the trait-level layer? if we introduce the concept of phoneme into this framework, we need to make assumption that phoneme boundaries are available . please note that 1/ you introduce a symbol and equation in this paper because you need them later. If you don't need them later, then you should not define such symbols, please check carefully all symbols in this paper.  )}
% {Despite much progress, most neural speaker models encode a spoken utterance into a speaker embedding. By comparing the speaker embeddings of an enrollment utterance and a test utterance,  we derive a score without providing insights into how the score is derived. To overcome this limitation, we introduce phonetic trait layers into the models. %to make users visualize and compare the speaker information in different phonemes. 

To explain how we derive the similarity score, {we introduce phonetic trait layers into speaker models. While the phonetic trait layers are designed for any speaker models. Without loss of generality, we take the commonly used ECAPA-TDNN~\cite{desplanques2020ecapa} as an example, which consists of frame-level and utterance-level layers. The trait layers seek to  generate phonetic traits from the frame sequence and pool these phonetic traits to obtain an utterance-level speaker embedding as in Fig.~\ref{fig:architecture} (a). This ensures that the output speaker embedding is derived from fine-to-coarse hierarchy explainable in human terms.}

%The architecture of our proposed model is illustrated in Fig.~\ref{fig:architecture}(a). 
%
%We introduced the trait-level layers in the green dash box within the commonly used black-box model.
%, which allows the model to capture the speaker information contained in each phoneme. %The speaker information contained in each phoneme is defined as the speaker attribute. 
%After that, phonetic traits are processed by the pooling layer to generate speaker embeddings. 
%The predicted phoneme sequences from a pre-trained phoneme classifier~\cite{zhu2022charsiu} are used to assign the phoneme label to each frame in both the training and testing phases. 
{We prepare the training and test utterances by applying a pre-trained wav2vec2-based phone recognizer~\cite{zhu2022charsiu}, that demarcates the phone boundaries.} This allows us to compute the phonetic traits in the `Trait Embedding' block as shown in Fig.~\ref{fig:architecture} (b). We first derive a frame-level embedding sequence of $T$ frames from the frame-level layers, where each embedding has $D_1$ dimensions. We then derive the phonetic trait by averaging the frame-level embeddings within each phone segment. Suppose we have a phonetic inventory of $I$ phones, the phonetic trait is represented by $I$ $D_1$-dimensional trait embeddings, which forms a \textit{phonetic trait set}  to characterize a speaker. %. This phonetic trait set is involved in model optimization, and a specific loss function will be introduced in Sec.\ref{sec3}
%The phonetic traits are computed in the ``Trait Embedding'' block as the average of frames belonging to the same phoneme:
%\vspace{-0.2cm}
%\begin{equation}
%h_i=\frac{1}{T^i}\sum_k z^i_k,
%\vspace{-0.2cm}
%\end{equation}
%where $z^i_k$ is the $k$-th frame in the output of the last frame-level layer that is assigned to $i$-th phoneme. $T^i$ is the number of frames belonging to $i$-th phoneme and $h_i$ is the $i$-th phonetic trait describing the speaker characteristic contained in $i$-th phoneme. 
%The total 40 phonetic traits are concatenated:
%\begin{equation}
%\vspace{-0.1cm}
%H=[h_1, h_2, ..., h_{40}].
%\label{phone-emb}
%\end{equation}
%
%The simplified process flow is illustrated in Fig.~\ref{fig:architecture}(b). 
%For efficient optimization, the order of phonemes in $H$ is fixed and consistent for every utterance. 

Considering that some phones might be absent in an utterance, we set the empty phonetic trait as $\textbf{0}$, or a blank vector in the colored phonetic trait set in Fig.~\ref{fig:architecture} (b). To reduce sparsity, such empty phonetic trait is removed by a `Trait Filter' block before the utterance-level layers, leading to $N \leq I$ trait embeddings.  
A statistics pooling layer aggregates all $N$ trait embeddings and propagated through a linear layer. We obtain a speaker embedding  of $D_2$ dimensions as the output of the linear layer, following the same setting as in ECAPA-TDNN.
 
%\textcolor{blue}{(we apply a pooling layer to handle variable $N$. In the next section, the loss functions are all applied to the phonetic trait set (which has a fixed shape of $D_1 \times I$). Maybe after introducing the phonetic trait set in the third paragraph, we emphasize that using it to optimize the network can reduce confusion.)} \textcolor{red}{in your blue comments, you first said variable N, then you said fixed shape $D_1 \times N$? is N fixed or variable? I don't know whether in Section III we are comparing $D_1 \times N$ or $D_2 \times 1 $? in Section III, it seems that we are comparing $D_1 \times I$! this is totally different from what we discuss around N in Section II. The symbol N actually disappears after Section II.} \textcolor{blue}{Sorry Prof it's a typo. the fixed shape is $D_1 \times I$. In Section~\ref{sec3} we are comparing $D_1 \times I$. The $N$ with different subscripts appears in the following sections, like $N_{e,k}$ in Eq.(2).}
\vspace{-0.3cm}
\section{Training and Loss Function}
\label{sec3}
%Usually, the speaker network is trained with a classification task, while is tested with a verification task. This mismatch may increase the opaqueness of the model. Therefore, apart from the traditional speaker classification task, we also trained the model with phoneme-dependent loss to simulate the speaker verification process. 
%\subsection{Overview} 

During training, each minibatch contains \emph{K} randomly selected speakers.  Each speaker contributes two randomly selected utterances: one for enrollment and  {another for test}. %\textcolor{red}{(why testing here during training? in machine learning, we have training, validation, test set that are clearly defined. what is the testing utterance here? - Haizhou)} \textcolor{blue}{We are only simulating the verification process during training with one enrollment utterance and one test utterance. If ``testing'' is likely to cause ambiguity, can we use ``query'' or ``probe'' instead? }
For simplicity, as shown in Fig.~\ref{fig:architecture}, we denote the set of enrollment utterances within a minibatch as \( \mathcal{U}_e = \{u_{e,1}, u_{e,2}, \ldots, u_{e,K}\} \) and the set of {test} utterances as \( \mathcal{U}_t  = \{u_{t,1}, u_{t,2}, \ldots, u_{t,K}\} \). 

An enrollment utterance \( u_{e,k} \) and a test utterance \( u_{t,k} \) from the same speaker form a matched pair \((u_{e,k}, u_{t,k})\).
Meanwhile, a randomly sampled utterance \( u_{t,h} \) (where \( h \neq k \)) from a different speaker is selected to form an unmatched pair \((u_{e,k}, u_{t,h})\). We further denote the fine-grained phonetic trait set \( e_k = \{e_k^1, e_k^2, \ldots, e_k^I\} \) for the enrollment utterance $u_{e,k}$ and \( t_k = \{t_k^1, t_k^2, \ldots, t_k^I\} \) for the test utterance $u_{t,k}$.
\vspace{-0.3cm}
\subsection{Trait verification loss}
%To maintain trait-level consistency for the same speaker and ensure differentiation across different speakers, we define the contrastive-style verification loss function as follows:
Usually, the speaker network is trained with a classification task, while is tested with a verification task~\cite{zheng2020phonetically,10094954,bai2022end}. This mismatch may increase the opaqueness of the model. To reduce this mismatch, we define a verification loss function $\mathcal{L}_{veri}$, that strengthens the trait consistency for the same speaker while discriminates the traits between speakers, %$\mathcal{L}_{same}$ for the matched pairs and  $\mathcal{L}_{diff}$ for unmatched pairs.  % We refer to these two losses collectively as specific losses ($\mathcal{L}_{spcf}$) for convenience.
%apart from the traditional speaker classification task, we also trained the model with phoneme-dependent loss to simulate the speaker verification process. 
\begin{equation}
\label{veri}
\begin{split}
\mathcal{L}_{veri}&= \frac{\alpha }{N_1} \sum_{i=1}^I \sum_{k=1}^{K} \mathbb{I}( e_k^i \neq \mathbf{0} \land t_k^i \neq \mathbf{0}) || e_k^i - t_k^i ||^2 \\
&-\frac{\beta}{N_2} \sum_{i=1}^I \sum_{k=1}^{K}  \min_{\substack{ h \neq k}} \mathbb{I}(e_k^i \neq \mathbf{0} \land t_h^i \neq \mathbf{0}) || e_k^i - t_h^i ||^2,
\end{split}
\end{equation}
where \(\mathbb{I}(\cdot)\) is the indicator function, having 1 if the condition is true and 0 otherwise. The symbol $\land$ represents the logical AND operator. \(N_1 = \sum_{i=1}^I \sum_{k=1}^{K} \mathbb{I}(e_k^i \neq \mathbf{0} \land t_k^i \neq \mathbf{0})\) and \(N_2 = \sum_{i=1}^I \sum_{k=1}^{K} \mathbb{I}(e_k^i \neq \mathbf{0} \land t_h^i \neq \mathbf{0})\) are two denominators. $\alpha$ and $\beta$ are hyper-parameters to balance the inter-speaker and intra-speaker trait loss. The least distance is retained among unmatched pairs, representing the most similar pair.%\textcolor{red}{(can we swap I and K summation to be consistent with Eq (2)? - Haizhou)}\textcolor{blue}{it is consistent. We can swap to make the order same with Eq (1)}
\vspace{-0.3cm}
\subsection{Trait-center loss}
%\textcolor{red}{(To Ma Yi: this loss is basically to minimize the variance of the trait vectors. I don't understand why. By minimizing the variance, we actually reduce the discriminative ability of the traits!)}
%\textcolor{blue}{(It's because the fact that not all phonemes have the same level of discrimination. During training, some traits are difficult to optimize with  $\mathcal{L}_{same}$ and $\mathcal{L}_{diff}$. For these traits, we use a more relaxed constraint (i.e. $\mathcal{L}_{center}$), aiming for them to become closer to each other. During training, we cannot determine which traits are discriminative enough for  $\mathcal{L}_{same}$ and $\mathcal{L}_{diff}$ and which are not. We can only balance the two losses by adjusting the weight of each loss. Our experimental results also confirm that using both losses simultaneously achieves better results than using either loss alone.)}
%The previous loss function defined in Eq.~\ref{veri} only considers ensuring within-trait consistency, while the relationships across different traits from the same speaker are neglected.
The trait verification loss $\mathcal{L}_{veri}$ only considers relationships involving the same phone from either the same or different speakers, without looking at different phones.
In text-independent speaker verification, it is desirable that we compare speech samples between different phones. It is expected that the phonetic traits from one single utterance are close to another. To address this, we introduce a trait center loss $\mathcal{L}_{center}$ for it: %Furthermore, given that speaker information is not uniformly distributed across phonemes~\cite{7846267, magrin2024effect, li2024phonemes}, some phonetic traits may lack discriminative to be effectively optimized by $\mathcal{L}{veri}$. Therefore, $\mathcal{L}_{center}$ optimize these traits as a relaxed constraint. The mathematical expression is:
 %So we design the trait-center loss $\mathcal{L}_{center}$ to %guarantee phonetic traits from an utterance are close to each other:
%make the phonetic traits from an utterance closer together:
%Despite some phonemes (e.g. [\textipa{N}]) containing rich speaker information for us to recognize the speaker using them, others may not be so discriminative. Therefore, 
%Assuming the average of all speaker attributes as the speaker prototype in phoneme-level, we design the phoneme-center loss to guarantee all speaker attributes are close to the speaker prototype:
\vspace{-0.2cm}

\begin{equation}
\begin{split}
\label{independent}
%, \mathcal{P}_j\neq\textbf{0}
\mathcal{L}_{center}&=\frac{\gamma}{\sum_{k=1}^{K} N_{e,k}}\sum_{i=1}^{I}\sum^K\limits_{k=1}\mathbb{I}(e_k^i \neq \mathbf{0})  ||e^i_k-\hat{e}_k||^2\\
&+\frac{\gamma}{\sum_{k=1}^{K} N_{t,k}}\sum_{i=1}^{I}\sum^K\limits_{k=1} \mathbb{I}(t_k^i \neq \mathbf{0}) ||t^i_k-\hat{t}_k||^2,
\vspace{-0.3cm}
\end{split}
\end{equation}
\begin{equation}
\hat{e}_k=\frac{1}{N_{e,k}} \sum^I\limits_{i=1} \mathbb{I}(e_k^i \neq \mathbf{0}) e^i_k,
\end{equation}
\begin{equation}
\hat{t}_k=\frac{1}{N_{t,k}} \sum^I\limits_{i=1}\mathbb{I}(t_k^i \neq \mathbf{0}) t^i_k,
\end{equation}
where $N_{e,k}=\sum_{i=1}^I \mathbb{I}(e_k^i \neq \mathbf{0})$ and $N_{t,k}=\sum_{i=1}^I \mathbb{I}(t_k^i \neq \mathbf{0})$ and $\gamma$ is the weight for $\mathcal{L}_{center}$.
%$\hat{\mathcal{E}}_j$ and $\hat{\mathcal{T}}_j$ are computed as the average of phonetic traits of the utterance $j$-th enrollment and test tutterance.

%The distance of a valid phonetic trait from $\hat{\mathcal{H}}_j$ is used as the phoneme-center loss.
%The total number of valid phonetic traits for utterance $j$ is $N_j$. 
\vspace{-0.3cm}
\subsection{Loss function}
The final loss function $\mathcal{L}$ for ExPO optimization is: 
\begin{equation}
    \label{loss}
    \vspace{-0.1cm}
    \mathcal{L_{all}}=\mathcal{L}_{AAM}+\mathcal{L}_{veri}+\mathcal{L}_{center}.
\end{equation}
%
%where  $\gamma$ is the weight for $\mathcal{L}_{center}$.
Besides the $\mathcal{L}_{veri}$ and $\mathcal{L}_{center}$, the Additive angular margin loss (AAM)-Softmax~\cite{deng2019arcface} $\mathcal{L}_{AAM}$ penalizes misclassification on the speaker identity. 
\vspace{-0.2cm}
\section{Experiments}
\label{sec4}
\subsection{Experiment setup}
We use VoxCeleb2~\cite{chung2018voxceleb2} for training. % and we evaluate our system on Vox1~\cite{}, SITW~\cite{} and Librispeech~\cite{}.
The hyperparameters are fine-tuned and decided based on Vox1-O, where $\alpha$, $\beta$ and $\gamma$ are set to 0.0007, 0.00001 and 0.0001, respectively. 
We evaluate the model on Vox1-H, Vox1-E, The speakers in
the wild (SITW)~\cite{mclaren16_interspeech}, and Librispeech~\cite{7178964}. We use the trial list of the core-core condition of SITW for both the development and evaluation parts. Since there is no official trial list for Librispeech, we randomly generated one target and one non-target trial for each utterance in the train-clean-100, train-clean-360, dev-clean, and test-clean subsets, resulting in a total of 275,752 trials.
%The phoneme sequence of both training and test datasets are generated by the pre-trained wav2vec2-based  phoneme recognition model~\cite{zhu2022charsiu}.
The phonetic inventory contains  39 units in the CMU phone set~\cite{phoset} and a non-verbal label [N-V], therefore, $I=40$. %, resulting in a total of 40 categories. 
Any frame that cannot be classified as a phoneme has been assigned to the non-verbal label.

The speaker verification system is an implementation of the WeSpeaker toolkit~\cite{wang2022wespeaker}. We follow the `ecapa-tdnn' configuration while setting the number of speakers in each minibatch as 32. $D_1$ and $D_2$ are set to 1,536 and 192 respectively. 
%Since only two utterances are selected for each speaker in each epoch, 
%We increase the number of epochs to 13,650 so that the number of samples used for training is approximately equal to the toolkit settings. 
We trained an ECAPA-TDNN as the baseline of the traditional black-box model.
Both the ECAPA-TDNN and ExPO are trained with the same formula. 
\begin{table*}[ht]
  %\vspace{-0.5cm}
  %\fontsize{9}{10}\selectfont
  
   \caption{A comparative study between ExPO and its baseline. The performance of the final score is denoted as `Final'; the evidence score as `Evd'. ExPO is trained with different loss functions listed in the `Model' column.  Trait Layer denotes the incorporation of   `Trait Embedding' and `Trait Filter'
 block in the model} 
  \label{eval}
  \centering
  \footnotesize
  \setlength{\tabcolsep}{0.6mm}{
\begin{tabular}{cccccccccccccccccc}
  \toprule
\multirow{2}{*}{Model} & \multicolumn{4}{c}{\underline{ \qquad \qquad Setting  \qquad  \qquad}} & \multirow{2}{*}{score} & \multicolumn{2}{c}{\underline{ \quad Vox1-O \quad}} & \multicolumn{2}{c}{\underline{\quad  Vox1-E \quad }}& \multicolumn{2}{c}{\underline{\quad  Vox1-H \quad }}& \multicolumn{2}{c}{\underline{\quad  SITW-dev \quad }} & \multicolumn{2}{c}{\underline{\quad  SITW-eval \quad }}& \multicolumn{2}{c}{\underline{\quad  Librispeech\quad }}\\

 %& \multirow{2}{*}{Pho Layer} & \multirow{2}{*}{$\alpha$} & \multirow{2}{*}{$\beta$} & \multirow{2}{*}{$\gamma$} & \multicolumn{2}{c}{\underline{\qquad All\qquad }} & \multicolumn{2}{c}{\underline{\qquad 5s\qquad }} & \multicolumn{2}{c}{\underline{\qquad 2s\qquad }} & \multicolumn{2}{c}{\underline{\qquad Diag\qquad }} & \multicolumn{2}{c}{\underline{\qquad Full\qquad }} \\  
 &\scriptsize Trait Layer & $\alpha$ & $\beta$ & $\gamma$ & &EER & minDCF & EER & minDCF & EER & minDCF & EER & minDCF & EER & minDCF & EER & minDCF \\ \hline
\multirow{2}{*}{ECAPA-TDNN} & \multirow{2}{*}{\XSolidBrush} & \multirow{2}{*}{0} & \multirow{2}{*}{0} & \multirow{2}{*}{0} &Final& \textbf{1.276} & \textbf{0.157} & \textbf{1.384} & \textbf{0.156} &\textbf{2.562} &\textbf{0.245} & \textbf{1.953} & \textbf{0.192} & 2.242 &\textbf{0.219}&\textbf{0.693}&\textbf{0.048} \\
 & &  &  &  &Evd& 21.23 & 0.967 & 21.84 & 0.990 & 30.98 & 0.999 & 21.73 & 0.947 & 23.84 & 0.974 &13.56&0.718\\
%  & &  &  && Full & 45.219 & 0.999 & 45.107 & 1.000 & 46.337 & 1.000 & 45.168 & 1.000 &45.317 & 0.999 & 42.040 & 1.000 \\
  \hline
  
\multirow{2}{*}{\makecell{ExPO\\($\mathcal{L}_{AAM}$)}}& \multirow{2}{*}{\Checkmark} & \multirow{2}{*}{0} & \multirow{2}{*}{0} & \multirow{2}{*}{0}&Final & 1.356 & 0.166 & 1.515 & 0.184 & 2.956 & 0.295 & 2.387 & 0.218& \textbf{2.132} & 0.223&0.848&0.060 \\
& &  &  &  &Evd& 7.831 & 0.707 & 8.123 & 0.786 & 14.708 & 0.913 & 9.280 & 0.676 & 10.663 & 0.744& 5.131 & 0.324  \\
%  & &  &  && Full & 30.789 & 0.998 & 31.731 &1.000 & 35.415 & 1.000 & 30.882 & 1.000 & 31.641 & 0.999&28.719&0.994 \\
  \hline
  
\multirow{2}{*}{\makecell{ExPO\\($\mathcal{L}_{AAM}$,$\mathcal{L}_{center}$)}} &\multirow{2}{*}{\Checkmark}  & \multirow{2}{*}{0} & \multirow{2}{*}{0}&\multirow{2}{*}{1e-4}&Final & 1.979 & 0.237 & 1.976 &0.209  & 3.677 & 0.316 & 2.772 & 0.242 & 3.226 & 0.281 &0.895 &0.063 \\
& &  &  &  &Evd& 13.64 & 0.914 & 13.86 & 0.947 & 23.82 & 0.982 & 16.98 & 0.883 &18.48 & 0.912&9.237&0.544 \\
%& &  &  && Full & 34.962 & 0.997 & 34.867 & 1.000 & 39.037 & 1.000 & 35.734 & 1.000 & 36.386 & 1.000&32.251& 0.989 \\
\hline

\multirow{2}{*}{\makecell{ExPO\\($\mathcal{L}_{AAM}$,$\mathcal{L}_{veri}$)}} &\multirow{2}{*}{\Checkmark}  & \multirow{2}{*}{7e-4} & \multirow{2}{*}{1e-5} & \multirow{2}{*}{0} &Final& 1.808 & 0.196 & 1.776 & 0.191 &3.000  & 0.282 & 2.888 & 0.242 & 3.034 & 0.293 &0.819 &0.057 \\
& &  &  &  &Evd& 13.29 & 0.866 & 13.46 & 0.879 & 20.62 & 0.938 & 14.893 & 0.785 & 16.37 & 0.843 &6.736&0.373 \\
%& &  &  && Full & 30.193 & 0.999& 29.375 & 1.000 & 31.976 & 1.000& 35.251 & 1.000 & 36.448 & 0.999 &20.898&0.988\\
\hline

\multirow{2}{*}{\makecell{ExPO\\($\mathcal{L}$)}}  & \multirow{2}{*}{\Checkmark} & \multirow{2}{*}{7e-4} & \multirow{2}{*}{1e-5} & \multirow{2}{*}{1e-4} & Final& 1.552 & 0.184 & 1.637 & 0.187 & 3.007 & 0.287& 2.503 & 0.237 & 2.597 & 0.262&0.804&0.060\\
& &  &  &  &Evd& \textbf{6.781} & \textbf{0.674} & \textbf{7.269} & \textbf{0.684} & \textbf{12.602} & \textbf{0.816} & \textbf{7.908} & \textbf{0.588} & \textbf{9.408} & \textbf{0.660} &\textbf{3.916}&\textbf{0.245} \\
%& &  &  && Full & \textbf{10.585} & \textbf{0.834} & \textbf{10.466} & \textbf{0.820} &\textbf{17.886} & \textbf{0.970} & \textbf{10.566} & \textbf{0.997} & \textbf{11.864} & \textbf{0.999} &\textbf{6.768}&\textbf{0.566}\\
\bottomrule
\vspace{-0.7cm}
\end{tabular}
}
\end{table*}

\vspace{-0.3cm}
\subsection{Evaluation}
For the speaker verification decision, the cosine similarity between two speaker embeddings, namely enrollment and test utterance, is adopted as the \textit{final score}, with EER and minDCF serving as the evaluation metrics.  %\textcolor{red}{(where do you define final score, and evidence score? If their difference is just the diagonal vs all elements? please confirm if the final score here is just the difference between two embeddings as indicated in Fig. 2 (a)?, )}
%\textcolor{blue}{(I have highlighted the definitions of the final score and evidence score in blue. The final score is the cosine similarity of the speaker embeddings in Fig. 2(a), which is a scalar. The evidence score is the average of the phonetic trait similarity matrix's diagonal, which is calculated from the phonetic trait set in Fig 2 (a))}
{As in manual voice comparison, we attempt to explain the speaker verification results by showing the similarity between phonetic traits. } A positive correlation between this phonetic-level similarity and the final score serves as an evaluation criterion for explainability.

Specifically, for each trial, a $I$-dimensional phonetic trait similarity vector $\textbf{s}$ is generated, of which an element $s(i)$ represents the phone-wise cosine similarity between the $i$-th phonetic trait of the enrollment utterance and the test utterance. %\textcolor{red}{(how is this similarity matrix related to Figure 2.(c)? is the matrix used only in testing? not in the training? is similarity matrix presented in Figure 3? if yes, please explain in Figure 3 caption.)}
%\textcolor{blue}{%{Not the entire similarity matrix is used in training. For matched pairs, only the diagonal of the similarity matrix is involved. For multiple unmatched pairs of an enroll utterance, only the diagonal element with the minimum value in the similarity matrix is involved in the training. }
%Sorry Prof, I just find we only need the diagonal instead of the whole matrix. I will make change accordingly. I think $S$ still is only used in test. In training we test Euclidean distance of phonetic trait, while the $S$ is in test is cosine distance. But I redrew the Fig. 2(c), do you think it's better than previous?}
%\textcolor{red}{ (I am not clear what is the use of evidence score in the experiments. which decision is made according to evidence score?)}\textcolor{blue}{The evidence score is not used to make direct decisions, we want to use the evidence score to propose a quantitative metric for evaluating the reliability of the model's interpretation. The closer the performance of the evidence score is to the final score, the more reliable the interpretation.}
%

If the phone-wise similarity vector behaves similarly to the utterance-level final score, one is able to use the similarity vector to explain the final decision. {Motivated by this idea, we propose an \textit{evidence score} as the non-zero average of its values}, that is not used to make speaker verification decision, but rather as a quantitative metric for the above explainability. The formula of the evidence score is:
\begin{equation}
score=\frac{1}{N}\sum_i^I\mathbb{I}(e^i \neq \mathbf{0} \land t^i \neq \mathbf{0}) S(i),
\end{equation} 
where $e$ and $t$ is the enrollment and test utterance respectively, and $N=\sum_i^I \mathbb{I}(e^i \neq \mathbf{0} \land t^i \neq \mathbf{0})$.
The closer the evidence score is to the final score, the better the explainability is. 
% reliability of the model's interpretation. that reports the voice comparison of phonetic traits between the same phonemes. %. This score compares each phonetic trait only with those of the same phoneme.
%\vspace{-0.25cm}
%\begin{equation}
%\frac{1}{N_{Diag}}\sum_m^{40} S_{mm},
%\vspace{-0.25cm}
%\end{equation} 
%The \textit{evidence score} can also be calculated as the non-zero average of all the values in the similarity matrix. This score is derived from comparing each phonetic trait with those of all phonemes.
%\vspace{-0.25cm}
%\begin{equation}
% \frac{1}{N_{Full}}\sum_{m}^{40}\sum_{n}^{40}S_{mn},   
% \vspace{-0.25cm}
%\end{equation} 
%
A closer match between the evidence score and the final score suggests a better explanation of the decision. 
%In this study, we prioritize comparing speakers using the same phonemes, as this approach aligns more closely with human cognition. %However, in text-independent speaker verification, it is often inevitable to compare speaker information across different phonemes. Consequently, using different phonemes for comparison serves as an auxiliary explanation for the decisions made by our network.
\vspace{-0.3cm}
\subsection{Discrimination of phonetic traits}

The phonetic traits are not equally informative. We assess the
 discriminability of phonetic traits by  F-ratio~\cite{wolf1972efficient,foulkes2012forensic}, which compares between-speaker variation to within-speaker variation. This is a method commonly used to identify potentially useful features in forensic voice comparison.
% We apply the idea of F-ratio to test the discrimination of each phonetic trait in ExPO.

We randomly sample the cosine similarity of the same phonetic traits within target trials to reflect the within-speaker similarity,  %, i.e. $s(i)$, represents $i$-th phonetic trait's within-speaker similarity if $\textbold{s}$ is from a target trial,
with non-target trials for between-speaker similarity. To ensure robustness, we repeat this sampling process 500 times and average the results to obtain the final measure of similarity. %We randomly sampled 500 times for each phonetic trait from target trails. The average represents the within-speaker similarity for this trait. Similarly, each phonetic trait's between-speaker similarity is computed from 500 non-target trials. 
The ratio of the within-speaker similarity to between-speaker similarity indicates the phonetic trait's discrimination. The higher the ratio, the better the discriminability. The phone ‘ZH’ was excluded from this test because its occurrence in all test sets was significantly less than 500 times.
\vspace{-0.2cm}
\section{Results}
\label{sec5}
\begin{figure}[t]
  \centering
  \includegraphics[trim=0cm 0.7cm 0cm 0cm,width=\linewidth]{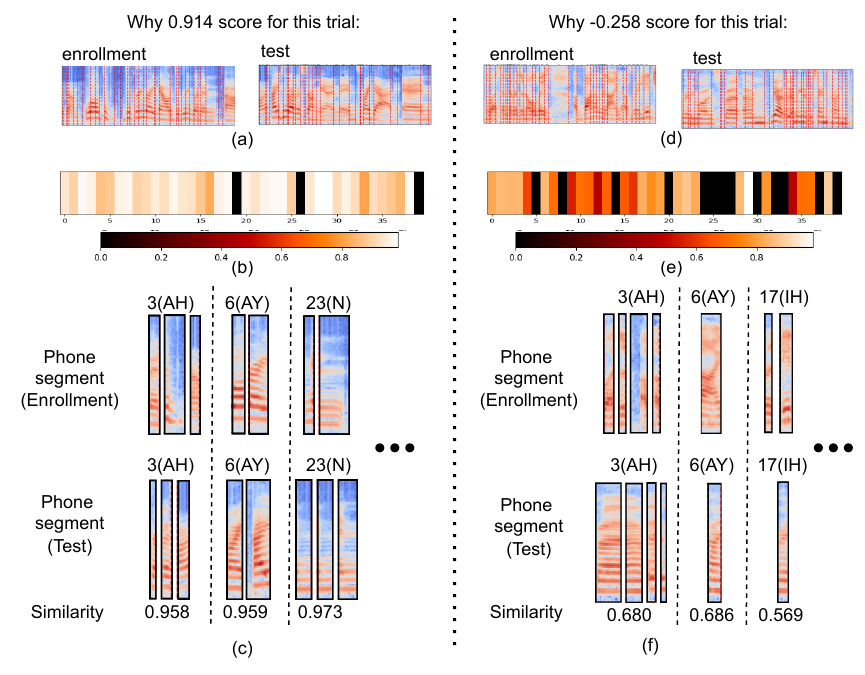}
  \vspace{-0.5cm}
  \caption{Visualization of the evidence provided by ExPO for two trials in Vox1-O. (a) and (d) are the spectrogram of enrollment and test utterances, where phone boundaries are marked by dotted lines. %. The red dashed line represents the phoneme boundary. %For clarity, only the first 3 seconds are illustrated. 
  {(b) and (e) are the phonetic trait similarity vector $\textbf{s}$ for (a) and (d) respectively.  $s(i)$ is the cosine similarity between ${i}$-th phonetic traits in enrollment and test utterance.} %Rows stand for the phoneme in the enrollment and columns is the phoneme in the test utterance. 
  The black bars indicate the absent phones in the utterances. (c) and (f) are visual comparisons of the spectrum, and the similarity of the phonetic traits.  }
\vspace{-0.5cm}
\label{evidence}
\end{figure}
\subsection{Explainable evidence}
Let us now understand the model's explainable results. Fig.~\ref{evidence} is a visualization for a positive and a negative trial. 

In the first trial, as in Fig. 3(a), the final score is 0.914, and the phonetic trait similarity vector $\textbf{s}$ in Fig. 3(b) supports this high score. 
In Fig.~\ref{evidence}(c), we visualize the phonetic trait in detail. For example, the first column of Fig.~\ref{evidence}(c) shows that the similarity of the third phonetic trait, i.e. phone `AH', of enrollment and test utterance is 0.958, as indicated by $s(3)$ in Fig. 3(b).  Other traits also show similar values, such as 0.959 for the 6th trait and 0.973 for the 23rd trait, further supporting the high score. Similarly, the right-hand side of Fig.~\ref{evidence} explains why the second trial resulted in a low final score of {-0.258}.
\begin{figure}[t]
  \centering
  \includegraphics[trim=0cm 0cm 0cm 0.8cm,width=0.45\textwidth]{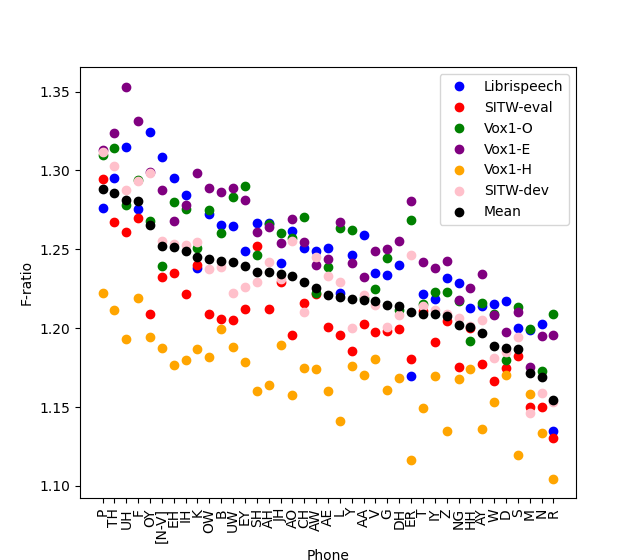}
 
  \caption{The F-ratio of phonetic trait extracted from different phones. The phones are ordered from highest to lowest based on their mean values across all test sets.}
\vspace{-0.5cm}
\label{discrimination}
\end{figure}
The phonetic trait matrix for this trial is shown in Fig.~\ref{evidence}(e). It indicates that the phonetic traits in this trial are less similar than those in Fig.~\ref{evidence}(a). Fig.~\ref{evidence}(f) provides a detailed comparison for this trial, revealing fewer similar traits than those on the left of Fig.~\ref{evidence}.
%Fig.~\ref{evidence} (e) and (f) explain why a lower final score is assigned to this trial.
\vspace{-0.4cm}
\subsection{Evaluation of accuracy and explainability}
In Table~\ref{eval}, we summarize the accuracy and explainability for the ECAPA-TDNN and ExPO. The ExPO trained with the loss function as formulated in Eq. (\ref{loss}), which minimizes the gap between the evidence scores (`Evd’) and the final score (`Final’), demonstrates the best explainability among all. We observe that the consistency between the performance of the final score and the evidence score is evident. In most cases, as the EER of the final score decreases, so does the EER of the evidence score. %This sugguests that ExPO accurately reflects the decisions made by the model.  

%We acknowledge that performance may slightly decline compared to the baseline. However, considering that the goal of our paper is transparent evidence, and given the inevitable trade-off between performance and interpretability, we believe this decline is acceptable. 
%We argue that this work focuses on providing transparent evidence rather than improving model performance. 
%Compared to existing interpretability methods that significantly sacrifice performance to achieve explainability, our proposed ExPO method demonstrates clear advantages. It provides better interpretability with only a slight reduction in performance, averaging just a 0.33\% increase in EER compared to the baseline. 
%Compared to other related work~\cite{wu2024explainable}, our model aligns the decision-making process with human perception and provides comprehensible evidence, while only averaging a 0.33\% increase in EER compared to the black-box model.
%This ensures ExPO's applicability in real-world scenarios and enhances the trustworthiness of deep learning systems. 
We emphasize that this work prioritizes explainability over enhancing model performance. Despite a slight 0.33\% increase in EER over that of the baseline model, this approach makes ExPO explainable in human terms. {Unlike other studies~\cite{wu2024explainable}, which sacrifice accuracy due to speaker attributes that lack sufficient discrimination, we maintain high accuracy in real-world scenarios, while simultaneously providing clear explanations.} %\textcolor{red}{(please be explicit about what [22] is doing?)}%our work  balance enhances the trustworthiness of deep learning systems in real-world scenarios.
%We argue that this work prioritizes explainability over enhancing model performance.  Our model aligns the decision-making process with human perception and offers comprehensible evidence, while only averaging a 0.33\% increase in EER compared to the baseline model. Compared to other related work~\cite{wu2024explainable}, this balance ensures ExPO's applicability in real-world scenarios and enhances the trustworthiness of deep learning systems.

Table~\ref{eval} presents an ablation study on the importance of using both $\mathcal{L}_{center}$ and $\mathcal{L}_{veri}$. Using either $\mathcal{L}_{center}$ or $\mathcal{L}_{veri}$ alone results in a decline in performance and explainability. This could be due to the fact that not all phones are equally distinctive. Traits with high distinctiveness are well-suited for $\mathcal{L}_{veri}$ optimization, while using only $\mathcal{L}_{veri}$ for traits with low distinctiveness may lead to overfitting or convergence issues. %$\mathcal{L}_{center}$ provides a more relaxed and reasonable constraint for these traits. %, indirectly indicating that the discrimination of each trait varies.%This confirms the hypothesis that these two losses complement each other. %Although the performance of our system drops slightly compared to the baseline, our target is to build a transparent speaker network. 
\vspace{-0.3cm}
\subsection{Discriminability of phonetic trait}
Fig.~\ref{discrimination} shows the F-ratio for the phonetic traits across multiple test sets. We observe that all phonetic traits achieve a ratio greater than 1.0,  confirming their discriminability. 
Although the ratio varies across different test sets, they generally follow the same trend decreasing from left to right in terms of F-ratio. This suggests that the discriminativeness of the phonetic traits is consistent across test sets. 
We observe that the `[N-V]' values are unexpectedly high, which suggests  that non-verbal segments may play a more substantial role than previously thought. This observation may be attributed to the potential speaker information in non-verbal voice, such as laughter, breathing, coughing, and also rhythmic speech patterns. The importance of non-verbal segments is also investigated in both automatic speaker verification~\cite{trouvain2014laughing,lin2023haha,janicki2012impact} and forensic speaker comparison~\cite{national1930theory,tiersma2012oxford}. 
We believe our study provides a cue for understanding the role of non-verbal sounds in speaker verification.

%Additionally, we found that nasal sounds generally have poor discriminability, which may differ from human perception in speaker recognition. %\textcolor{red}{(I am not sure if you have normalize the F-ratio for different duration of phonemes?)}\textcolor{blue}{(In our calculation of the F-ratio, we base it on phonetic traits, which is already averaged over frames. So I didn't further normalize for the duration of the phonemes.)}
%we observe that the ratios for most vowel phonemes' speaker attributes are higher than others, which aligns with the human perception that vowels are easier for speaker identification. 
%We hypothesis the high discrimination power of `[N-V]' stems from the inclusion of non-verbal sounds like laughter, breathing, and coughing, which provide speaker-specific acoustic cues useful for identity verification.
\vspace{-0.3cm}
 \section{Conclusion}
\label{sec6}
%In conclusion, while deep learning models have advanced speaker verification, achieving state-of-the-art performance, the lack of explainability remains a challenge. 
%
%Interpretability efforts have primarily focused on post-hoc explanations.
%or designing inherently interpretable models. 
%
%However, existing methods face limitations in practical application due to the reliance on fixed heat maps and the closed-set nature of training data. 
%
In this work, we validated an idea to integrate phonetic traits in a deep learning-based speaker network to enhance model explainability.
We shown that, by comparing phonetic traits between enrollment and test utterances, the proposed ExPO model offers a better understanding of the decision-making processes in a neural speaker verification system, aligning with human perception. The research marks a step towards explainable speaker verification.

\bibliography{ref.bib}
\bibliographystyle{IEEEtran}

\end{document}